\documentclass{iaus}
\usepackage{graphicx}

\title[Deep Impact ejection] 
{Deep Impact ejection from Comet 9P/Tempel 1 as a triggered outburst}

\author[Sergei I. Ipatov \& Michael F. A'Hearn]   
{Sergei I. Ipatov$^1$
 \and Michael F. A'Hearn$^2$}

\affiliation{$^1$Catholic University of America \\ Washington DC, USA \\ email: {\tt siipatov@hotmail.com} \\[\affilskip]
$^2$Dept. of Astronomy, University of Maryland, \\ College Park, MD, USA \\email: {\tt ma@astro.umd.edu}}

\pubyear{2010}
\volume{263}  
\pagerange{317--320}
\jname{Icy Bodies of the Solar System}
\editors{J.A. Fernandez, D. Lazzaro, D. Prialnik, \& R. Schulz, eds.}
\begin{document}

\maketitle

\begin{abstract}
Ejection of material after the Deep Impact collision with Comet
Tempel 1 was studied based on analysis of the images made by the Deep Impact
cameras during the first 13 minutes after 
impact.
Analysis of the images shows that there was a local maximum
of the rate of ejection at time of ejection $\sim$10 s
with typical velocities $\sim$100 m/s.
At the same time, a considerable excessive ejection in a few
directions began, the direction to the brightest pixel
changed by $\sim$50$^\circ$, and there was a local increase of
brightness of the brightest pixel. 
The ejection can be considered as a superposition of the normal
ejection and the longer triggered outburst.
\end{abstract}

\firstsection 
\section{Analysis of images}

In 2005 the Deep Impact (DI) impactor collided with Comet 9P/Tempel 1 (A'Hearn et al. 2005).  Our studies (Ipatov \& A'Hearn 2008, 2009) of the time variations in the projections $v_p$ of characteristic velocities of ejected material onto the plane perpendicular to the line of sight and of the relative rate $r_{te}$ of ejection were based on analysis of the images made by the DI cameras during the first 13 min after the impact.  We studied velocities of the particles that were the main contributors to the brightness of the cloud of ejected material, that is, mainly particles with diameter $d<3$ $\mu$m. 
Below we present a short description of  analysis of the images and the conclusions based on our studies. Details of the studies and more figures and references are presented in Ipatov \& A'Hearn (2008). 
More complicated models of ejection will be studied in future.


Several series of images  taken through a clear filter were analyzed. In each series,
the total integration time and the number of pixels in an image
were the same. As in other DI papers, original images were rotated by 90$^\circ$ in anti-clockwise direction. 
In DI images, calibrated physical surface brightness (hereafter CPSB, always in W m$^{-2}$ sterad$^{-1}$ micron$^{-1}$) is presented. 
For several series, we considered the differences in brightness between images made after impact and a corresponding image made just before impact. 
Overlapping of considered time intervals for different series of images allowed us to calculate the relative brightness at different times
$t$ after impact,
though because of non-ideal calibration, the values of peak brightness at almost the same time could be different (up to the factor of 1.6) for different series of images. 
Variation in brightness of the brightest pixel and the 
direction from the place of ejection to the brightest pixel were studied.
At 
$t$$>$$100$ s, some DI images do not allow one to find accurately the relative brightness of the brightest pixel and the direction from the place of ejection
to this pixel. First, there
are large regions of saturated pixels on DI images, which may not allow one to calculate accurately the peak brightness. 
Secondly, at $t$$\sim$600--800 s, coordinates of the brightest pixel were exactly the same for several images. It can mean that some pixels became `hot' when the distance between the spacecraft and the nucleus of the comet became small. 

\begin{figure}[]
\begin{center}
\includegraphics[width=5.0in]{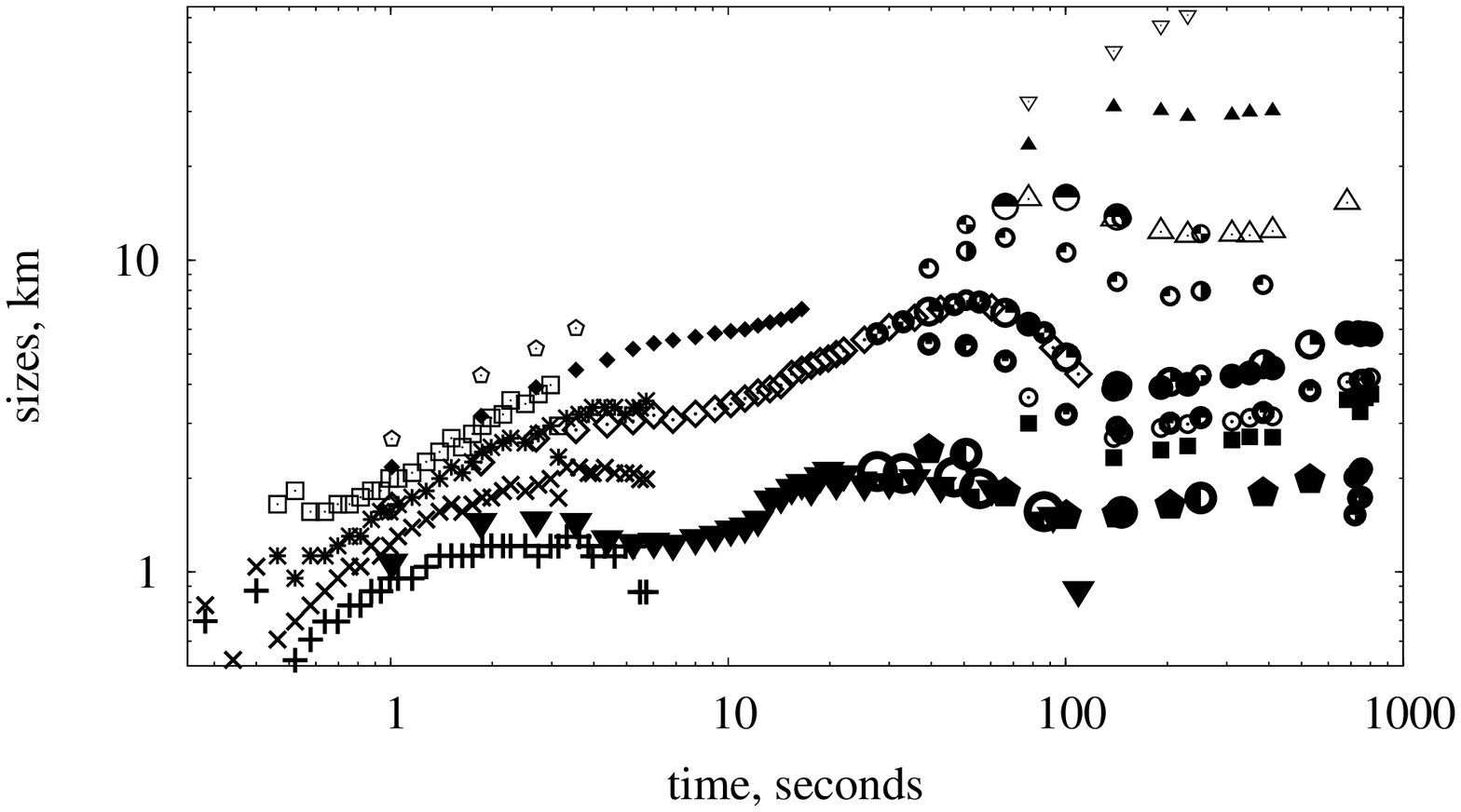}
\caption{Time variations in sizes $L$ (in km) of regions inside contours of CPSB=$C$. Different signs correspond to different series of images at different $C$. The curves have local minima and maxima that were used for analysis of time variations in velocities.}
\includegraphics[width=5.0in]{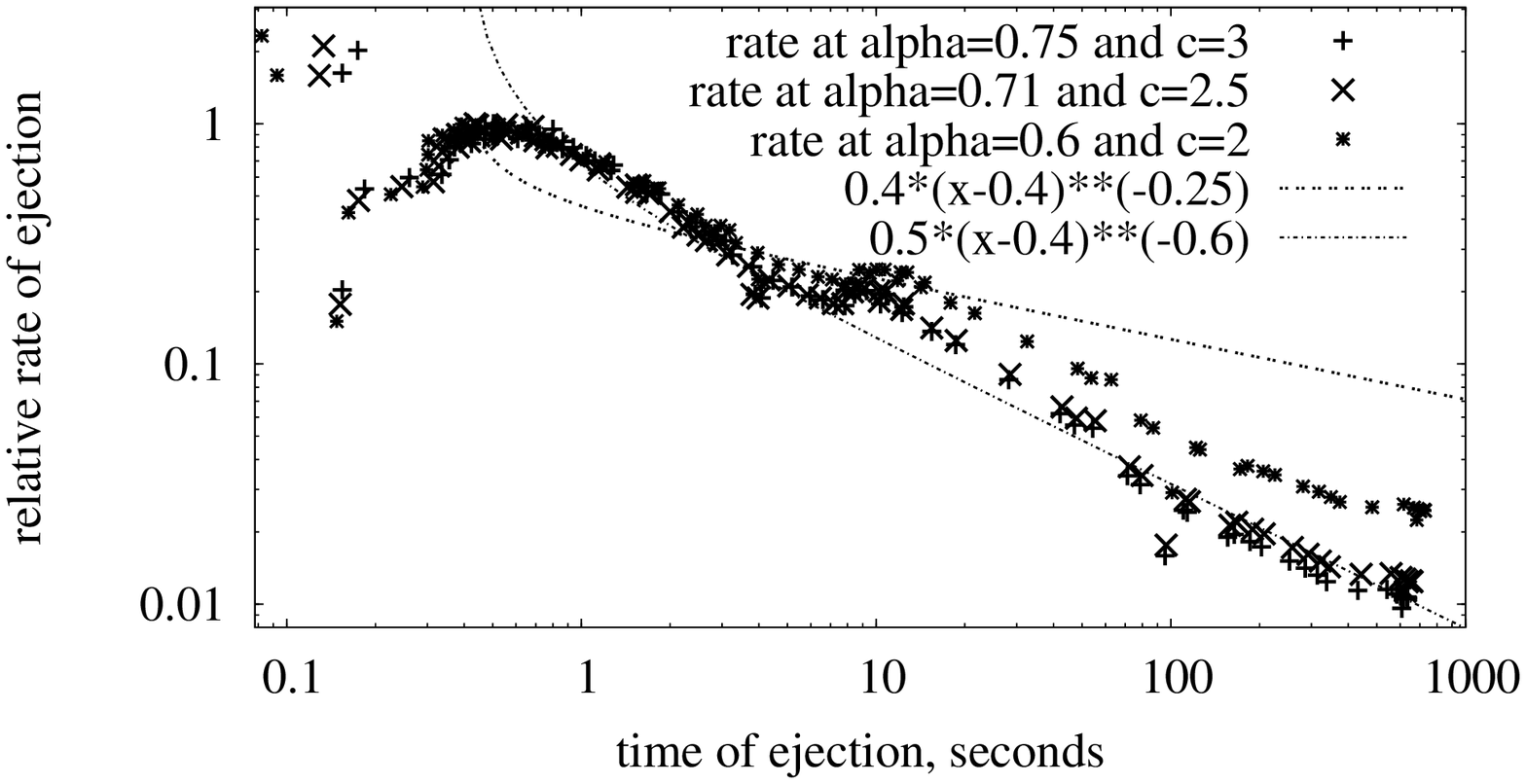}
\caption{Relative rate $r_{te}$ of ejection at different times $t_e$ of ejection for the model in which the characteristic 
velocities of the edge of the observed bright region at time $t$ are equal to $v_{expt}$=$c\cdot(t/0.26)^{-\alpha}$ (in km/s), for three pairs of $\alpha$ and $c$. 
The maximum rate at $t_e$$>$0.3 s is considered to be equal to 1. 
Two curves of the type $y$=$c_r\cdot(x-c_t)^{\beta}$ are also presented for comparison.}
\label{fig1}
\end{center}
\end{figure}

We analyzed the time dependencies of the distances $L$ from the place of ejection to contours of CPSB=const for several levels of brightness and different series of images (Fig. 1). Based on the supposition that the same moving particles corresponded to different local maxima (or minima) of $L(t)$ (e.g., to values $L_1$ and $L_2$ on images made at times $t_1$ and $t_2$), we calculated the characteristic velocities $v$=$(L_1$-$L_2$)/($t_1$-$t_2)$ at 
time of ejection $t_e$=$t_1$-$L_1/v$. 
In this case, we use results of studies of series of images in order to obtain one pair of $v_p$ and $t_e$. Such approach to calculations of velocities doesn't take into account that particles ejected at the same time could have different velocities.  
According to theoretical studies presented by Housen et al. (1983),
velocities of ejecta are proportional to $t_e^{-\alpha}$
(with $\alpha$ between 0.6 and 0.75), and
the rate $r_{te}$ of ejection is proportional to $t_e^{0.2}$ at $\alpha$=0.6 and to $t_e^{-0.25}$  at $\alpha$=0.75.
Our estimates of the pairs of $v_p$ and $t_e$ were compared
with the plots of $v_{expt}$=$v_p$=$c(t/0.26)^{-\alpha}$ at several pairs of $\alpha$ and $c$ (0.26 was considered because the second ejection began mainly at $t_e$$\approx$0.26 s). The comparison testifies in favor of mean values of $\alpha$ $\sim$0.7.
Destruction and sublimation of particles and variation in their temperature could affect on the brightness of the DI cloud,
but, in our opinion, don't affect considerably on our estimates of velocities
and slightly change estimates of $\alpha$.

For the edge of bright region (usually at CPSB=3), the values of $L$=$L_b$ in km at approximately the same time can be different for different series of DI images considered. Therefore, we cannot simply compare $L_b$ for different images, but need to calculate the relative characteristic
size $L_r$ of the bright region, which compensates non-ideal calibration of images and characterizes the size of the bright region. 
Considering that the time needed for particles to travel a distance $L_r$ is equal to $dt$=$L_r/v_{expt}$, we find the time $t_e$$=$$t$$-$$dt$ of ejection of material of the contour of the bright region considered at time $t$. 
The volume $V_{ol}$ of a spherical shell of radius $L_r$ and width $h$ is proportional to $L_r^2 h$, and the number of particles per unit of
volume is proportional to $r_{te} \cdot (V_{ol} \cdot v)^{-1}$, where $v$ is the velocity of the material. Here $r_{te}$ corresponds to the material
that was ejected at $t_e$ and reached the shell with $L_r$ at time $t$.
The number of particles on a line of sight, and so the brightness $B_r$, are approximately proportional to the number of particles per unit of volume
multiplied by the length of the segment of the line of sight inside the DI cloud, which is proportional to $L_r$. Actually, the line of sight crosses many shells characterized
by different $r_{te}$, but
as a first approximation we supposed that $B_r$$\propto$$r_{te} (v \cdot L_r)^{-1}$. For the edge of the bright region,  $B_r$$\approx$const.
 Considering $v$=$v_{expt}$,
we calculated the relative rate of ejection as $r_{te}$=$L_r t^{-\alpha}$.
Based on this dependence of $r_{te}$ on time $t$ and on the obtained relationship between $t$ and $t_e$, we constructed the plots of dependences of $r_{te}$ on $t_e$ (Fig. 2). 
Because of high temperature and brightness of ejecta, the real values
of $r_{te}$ at $t_e$$<$1 s are smaller than those in Fig. 2. 
As typical sizes of ejected particles increased with $t_e$, the real rate of ejection
decreased more slowly than the plot in this figure.
If, due to the outburst, typical velocities of observed ejected particles
did not decrease much at $t_e$$>$100 s, then the values of $r_{te}$
could be greater than those in Fig. 2.


Excessive ejection in several directions ('rays' of ejection) was studied based on analysis of the form of contours of constant brightness.
Bumps of the contours considered by Ipatov \& A'Hearn (2008) include the upper-right
bump ($\psi$$\sim$60-80$^\circ$, still seen at $t$$\sim$13 min), the right bump ($\psi$ increased from
90$^\circ$ at $t$$\sim$4-8 s to 110-120$^\circ$ at $t$$\sim$25-400 s), the
left bump ($\psi$$\sim$245-260$^\circ$), which transformed with time into the down-left
bump ($\psi$$\sim$210-235$^\circ$), the upper bump (backward ejection,
$\psi$ varied from 0 to -25$^\circ$, the bump consisted mainly of particles ejected after 80 s),
where $\psi$ is the angle between the upper direction and the direction 
to a bump measured in a clockwise direction.
Together with hydrodynamics of the explosion, icy conglomerates
of different sizes at different places of the ejected part of the comet
could affect on the formation of the rays.


\section{Conclusions}

Our studies showed that
there was a local maximum of the rate of ejection at 
time of ejection $t_e$$\sim$10 s (Fig. 2) with typical projections (onto the plane perpendicular to the line of sight) of velocities $v_p$$\sim$100--200 m/s. At the same time, the considerable excessive ejection in a few directions (rays of ejecta) began, the direction to the brightest pixel quickly changed by about 50$^{\circ}$, and there was a local increase of brightness of the brightest pixel. 
On images made during the first 10--12 s, the direction was mainly close to the direction of impact; after the jump, it gradually came closer to the direction of impact. 
These features at $t_e$$\sim$10 s were not predicted by theoretical models of ejection and could be caused by the triggered outburst.
At the  outburst, the ejection could be from entire surface of the crater,
while the normal ejection was mainly from its edges.
Starting from 10 s, the ejection of more icy material could begin. The increase in the fraction of icy material caused an increase in the observed ejection rate and the initial velocities (compared to the normal ejection). 

At $1$$<$$t_e$$<$3 s and $8$$<$$t_e$$<$$60$ s, the plot of time variation in estimated rate $r_{te}$ 
of ejection of observed material was essentially greater than the exponential line connecting
the values of $r_{te}$ at 1 and 300 s.
The difference could be mainly caused by that fact that the impact was a trigger of an outburst. 
The sharp decrease of the rate of ejection at $t_e$$\sim$$60$ s
could be caused  by the decrease of the outburst and/or of the normal ejection.
 The contribution of the outburst to the brightness of the cloud could be considerable, but its contribution to the ejected mass could be relatively small. 
 Duration of the outburst (up to 30--60 min) could be longer than that of the normal ejection (a few minutes). 
The studies testify in favor of a model close to gravity-dominated cratering.


Projections $v_p$ of the velocities of most of the observed material ejected at $t_e$$\sim$$0.2$ s were about 7 km/s. As the first approximation, the time variations in characteristic velocity at 1$<$$t_e$$<$100 s can be considered to be proportional to $t_e^{-0.75}$ or $t_e^{-0.7}$, but they could differ from this exponential dependence. The fractions of observed small particles ejected (at $t_e$$\le$6 s and $t_e$$\le$14 s) with $v_p$$>$200 m/s and $v_p$$>$100 m/s were estimated  to be about 0.13--0.15 and 0.22--0.25, respectively, if we consider only material observed during the first 13 min and $\alpha$$\sim$0.7--0.75. These estimates are in accordance with the previous estimates (100--200 m/s) of the projection of the velocity of the leading edge of the DI dust cloud, based on various ground-based observations and observations made by space telescopes. 
The fraction of observed material ejected with velocities greater than
100 m/s was greater than the estimates based on 
experiments and theoretical models. Holsapple \& Housen (2007) concluded that the 
increase of velocities
was caused by vaporization of ice in the plume and by fast moving gas.
In our opinion, the greater role in 
the increase of high-velocity ejecta
could be played by the outburst 
(by the increase of ejection of bright particles), and it may be possible to consider the ejection as a superposition of the normal ejection and the longer triggered outburst. 
Time variations in velocities could be smaller (especially, at $t_e$$>$100 s) for the outburst ejecta than for the normal ejecta.

The excess ejection of material in a few directions (rays of ejected material) was considerable during the first 100 s
 and was still observed in images at $t$$\sim$500--770 s. 
The sharpest rays were caused by material ejected at $\sim$20 s. 
 In particular, there were excessive ejections, especially in images at $t$$\sim$25--50 s, in directions perpendicular to the direction of impact. Directions of excessive ejection could vary with time. 

The work was supported by NASA DDAP grant NNX08AG25G.

\end{document}